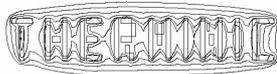

# Thermal and Mechanical Analysis of High-Power Light-Emitting Diodes with Ceramic Packages

Jianzheng Hu, Lianqiao Yang, and Moo Whan Shin

Department of Materials Science and Engineering, Myongji University, 38-2 Nam-Dong, Yonin-Si, Kyunggi-Do, 449-728, Korea

mwshin@mju.ac.kr

*Abstract*-In this paper we present the thermal and mechanical analysis of high-power light-emitting diodes (LEDs) with ceramic packages. Transient thermal measurements and thermo-mechanical simulation were performed to study the thermal and mechanical characteristics of ceramic packages. Thermal resistance from the junction to the ambient was decreased from 76.1 °C/W to 45.3 °C/W by replacing plastic mould to ceramic mould for LED packages. Higher level of thermo-mechanical stresses in the chip were found for LEDs with ceramic packages despite of less mismatching coefficients of thermal expansion comparing with plastic packages. The results suggest that the thermal performance of LEDs can be improved by using ceramic packages, but the mounting process of the high power LEDs with ceramic packages is critically important and should be in charge of delaminating interface layers in the packages.

## I. INTRODUCTION

The research efforts on GaN-based light-emitting diodes (LEDs) are kept increasing due to their significant impacts on illumination industry. One of key factors possibly limiting them from such an important application is the life time of LED package. The junction temperature of LEDs is greatly dependent on the thermal performance of LED packages and is believed to have great effect on the reliability of LEDs [1]. Further more, LED packages have to experience several thermal cycles during the surface mounting process. Delamination is possibly induced in the mounting process and directly responsible for mechanical reliability problems [2, 3]. A good package design by using proper materials and structures is important to ensure both the thermal and mechanical reliability of LED packages.

Of particular interests are ceramic materials for LED packages, which have been widely used in modern electronic packaging because of various advantages over plastic materials. The advantages include excellent thermal conductivity, excellent endurance to withstand heat, ability to withstand hazardous environments, flexibility for small and thin structures, enhanced reflectivity due to advanced surface finishing technology, less coefficient of thermal expansion (CTE) mismatch with the die, and high moisture-proof ability [4, 5]. In particular, excellent thermal characteristics of ceramic materials made them one of prime candidates for high power operation [6, 7]. Despite of the advantages of ceramic material as a packaging mold for LEDs, there still have not been found sufficient reports on thermal and mechanical analysis of ceramic LED packages.

This paper reports on thermal and mechanical performances of high power LEDs packaged with $Al_2O_3$-based ceramic mold. Transient thermal measurements were performed to characterize the thermal performance of ceramic packages. And coupled thermal and mechanical analysis was realized by FEA to study the mechanical performance.

## II. EXPERIMENT AND SIMULATION

LED samples with ceramic package were fabricated according to following process. Ceramic sheets were blanked and then punched in order to form via holes. The electrical circuits were patterned and coated with silver. The next steps are layer lamination, cutting, firing, and breaking into separate ones. The final steps are locating the die attachment and chip, die bonding, epoxy lens casting, and curing the epoxy. Transient thermal measurements with a Transient Thermal Tester (T3Ster, MicReD Ltd.) were performed to investigate the thermal behavior of the LEDs with ceramic packages. T3Ster captures the thermal transients real-timely, records the cooling /heating curve and then evaluate the cooling / heating curve to derive the thermal characteristics. Based on the thermal R-C network and structure function theory, the heat conducting path can be determined quantitatively. The measurements were done in a liquid bath (Julabo FP50) according to the following procedures. The first step was to get the K factor, a constant defining the relationship between the junction temperature change and temperature sensitive parameter (TSP). For compounds LEDs, TSP is the forward voltage drop of the diode. In the K factor calibration, 10 mA sensor current was used in the temperature range of 20 °C to 50 °C with an increasing step of 10 °C. Transient measurement was started to record the cooling curve after driving the samples with 350 mA current for 10 minutes at the liquid temperature of 25 °C to reach the thermal stabilization. The mechanical performance was evaluated by simulating the heat block test for electronic packages. In the heat block test, the packaged samples are heated in the isothermal heat block of 270 °C for 20 seconds. The transient thermal analysis followed by the structural analysis was done by commercial FEA code (ANSYS V9.0). The obtained thermo-mechanical stress induced in the heat block test was used to characterize the mechanical performance of LEDs with ceramic packages.

III. MODELING FOR THE MEASUREMENT AND SIMULATION

A. *Modeling for the Transient Thermal Measurement*

In the transient thermal measurement, a thermal step function response as either heating or cooling curves was provided by T3Ster. Based on the characteristics of LEDs, cooling curves were used in our measurement. The cooling curves were evaluated to extract the structure functions. The differential structure function [8, 9] is defined as the derivative of the cumulative thermal capacitance with respect to the cumulative thermal resistance and is expressed in (1). Equation (1) can be further transformed to (2) by the definition. In (1) and (2), $c$ is the volumetric heat capacitance, $\lambda$ is the thermal conductivity, and $A$ is the cross sectional area of the heat flow. So the thermal resistance change caused by $c$, $\lambda$ or $A$ can be observed in the plot of the differential structure function [10].

$$K(R_\Sigma) = \frac{dC_\Sigma}{dR_\Sigma} \quad (1)$$

$$K(R_\Sigma) = \frac{cAdx}{dx/\lambda A} = c\lambda A^2 \quad (2)$$

B. *Thermal Modeling*

The transient heat conduction equation without heat generation shown in (3) was used to obtain the temperature distribution during the heat block test.

$$\frac{\partial^2 T}{\partial x^2} + \frac{\partial^2 T}{\partial y^2} + \frac{\partial^2 T}{\partial z^2} = \frac{1}{\alpha}\frac{\partial T}{\partial t} \quad (3)$$

where the property $\alpha = k/\rho C$ is thermal diffusivity of the material, $k$ is the thermal conductivity, $\rho$ is the density, and $C$ is the specific heat. $T$ is the temperature, $x$, $y$ and $z$ are the spatial coordinates. For the heat block test, the boundary condition is to apply heat transfer coefficient combining convection and radiation heat transfer to the external surfaces of the LED package. The whole heat transfer coefficient varies with the temperatures of different surfaces of the package, and the ambient temperature is fixed at 270 ˚C. The thermal parameters of LED packaging materials are listed in TABLE I.

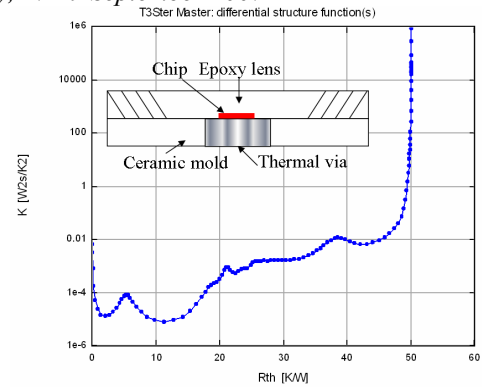

Fig. 1. Differential structure function of LEDs with ceramic package; the inset graph is the schematic of the ceramic package structure.

The modeled LED package has a dimension of 5.06×5.06×0.95 mm$^3$ and the package is composed of chip, heat slug, lead frame, ceramic mould, and epoxy lens as shown in Fig. 1. The investigated LED samples are non-flip chip, so the sapphire substrate is attached to the heat slug by silver paste in the model. And the chip was modeled as sapphire for the material properties. The detailed structure of epi-layers in the chip was simplified and the effect of gold wire was not considered in modeling. The silver paste was used for the die attachment. Silver was used for the heat slug and lead frame, ceramic for the mould, and epoxy for the lens. As the structure of the samples is symmetrical, only half of the structure was modeled in the simulation to save computing resources and improve efficiency. All the structural parameters employed in the simulation have the same values as in the real LED package investigated.

C. *Thermo-mechanical Modeling*

As shown in (4), thermal stress comes from the expansion of material when being heated.

$$\sigma = E\alpha(T - T_{ref}) \quad (4)$$

where the constant, $E$, is elastic modulus. $\alpha$, $T$, and $T_{ref}$ is CTE, temperature and reference temperature, respectively.

TABLE I
THERMAL PROPERTY OF LED PACKAGING MATERIALS

| Material | sapphire | die attach | silver | mould | | Epoxy lens |
| --- | --- | --- | --- | --- | --- | --- |
| | | | | ceramic | plastic | |
| $k$ (W/M·°C) | 35.1 | 7.5 | 419 | 3.2 | 0.23 | 0.17 |
| $C$ (J/Kg·°C) | 753.12 | 300 | 234 | 989 | 1256 | 1173 |
| $\rho$ (Kg/m$^3$) | 3980 | 2400 | 10491 | 3100 | 1300 | 980 |

TABLE II
MECHANICAL PROPERTY OF LED PACKAGING MATERIALS

| Material | sapphire | die attach | silver | mould | | epoxy lens |
| --- | --- | --- | --- | --- | --- | --- |
| | | | | ceramic | plastic | |
| $\alpha$, CTE (10$^{-6}$/K) | 5.6 | 30 | 19.9 | 5.8 | 45 | 45 |
| Poison's ratio | 0.25 | 0.35 | 0.39 | 0.24 | 0.35 | 0.49 |
| $E$ (GPa) | 335 | 40 | 76 | 76 | 5.2 | 0.5 |

The temperature distribution obtained from the transient thermal analysis for the heat block test is applied as temperature loads for the model. The reference temperature was set to the initial temperature 25 $^o$C. The mechanical properties of materials are temperature dependent, and the mean values for them are listed in Table II. The bottom of LED package is simply supported to limit the rigid displacement.

## IV. RESULTS AND DISCUSSIONS

### A. Thermal Measurement

K factor was calculated based on (5). The K factor calibrated for the measurement is 1.543 mV/$^o$C as shown in Fig. 2.

$$K = \Delta T_J / \Delta V_F \quad (5)$$

where $\Delta T_J$ is the change of the junction temperature of LEDs, and $\Delta V_F$ is the change of the forward voltage in LEDs.

The structure function obtained from the measurement is shown in Fig. 1. The thermal resistance from the junction to the ambient is 48.9 $^o$C/W. A numerical model built with finite volume method (FVM) was verified by the measured data, and

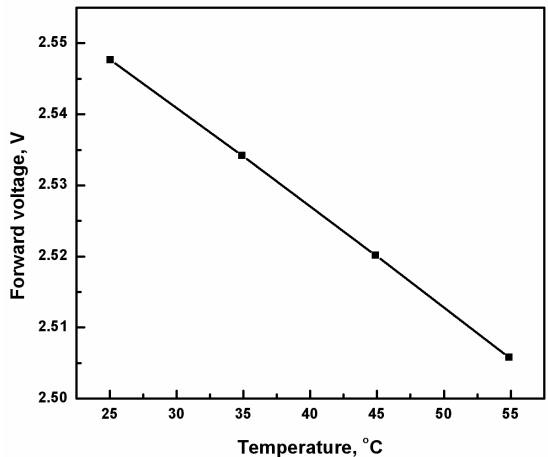
Fig. 2. Forward voltage vs. temperature plot for the calculation of the K factor.

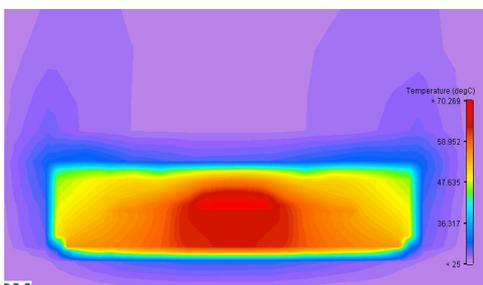
Fig. 3. Simulated temperature distribution in the ceramic package: the calculation was made in Flotherm.

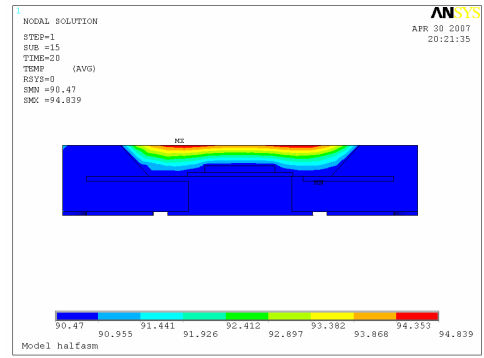
(a)

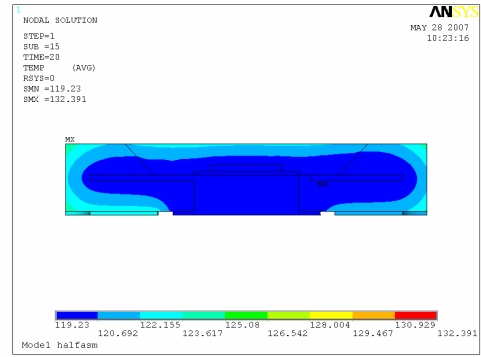
(b)

Fig. 4. Temperature distributions of LED packages with (a) ceramic mould and (b) plastic mould, respectively: the calculation was made by FEA.

then used to project the thermal resistance of LED package with plastic mould instead of ceramic mould. The calculated junction temperature distribution of the LEDs with ceramic packages was plot in Fig. 3. The thermal resistance from the junction to the ambient calculated from the simulation is 45.3 $^o$C/W for LEDs with ceramic packages. Using the same numerical model, by changing the material of mould from ceramic to plastic, thermal resistance of the LED package with plastic mould was obtained to be 76.1 $^o$C/W. So the thermal performance improving was realized by decreasing the thermal resistance of about 30 $^o$C/W comparing with the plastic packages.

### B. Thermal and Thermo-mechanical Simulation

In the transient thermal analysis it was observed that the temperature of chip increases from initial temperature 25˚C to about 95˚C in the 20 seconds heating process. The large temperature gradient mainly existed between the epoxy lens and other package components as shown in Fig. 4(a). The temperature gradient among the chip, heat slug and ceramic mould is low because of higher thermal conductivity of these components. Compared with the ceramic package, the higher junction temperature and temperature gradient in the package were obtained because of the poor thermal conductibility of plastic as shown in Fig. 4(b).

The temperature gradient and the mismatching CTEs have induced high thermo-mechanical stress in the packages. The calculated thermo-mechanical stress in the ceramic package was shown in Fig. 5. As shown in Fig. 5 high stress existed in the chip and die attachment. The interface areas, especially the outer edge, showed higher stress level. The maximum Von. Mises stress is 931 MPa and 211 MPa for the chip and silver epoxy die attachment, respectively. The maximum value of the stress appeared at the corners of the chip and die attachment. This stress can lead to interface delaminations between the chip and die attachment. The delamination begins from the corners and then expands to other areas.

As a comparison, the thermo-mechanical stress in the LED package with plastic mould was also calculated and plotted in Fig. 6. The maximum Von. Mises stress is 751 MPa and 193 MPa for the chip and silver epoxy die attachment, respectively. Despite of much less CTE mismatching in the packages with ceramic mould, the thermo-mechanical stress level is higher than the packages with plastic mould. The large difference between the elastic moduli of the ceramic and plastic was proposed to be responsible. As the ceramic is much harder than the plastic, higher stress was induced in the chip even the CTEs are less mismatching.

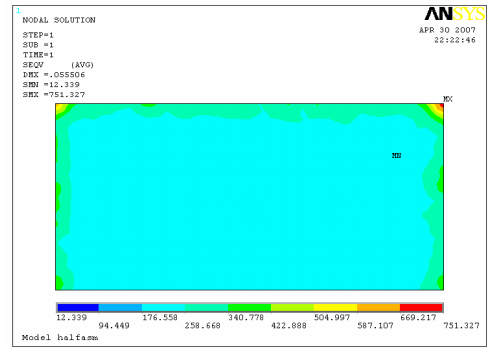

(a)

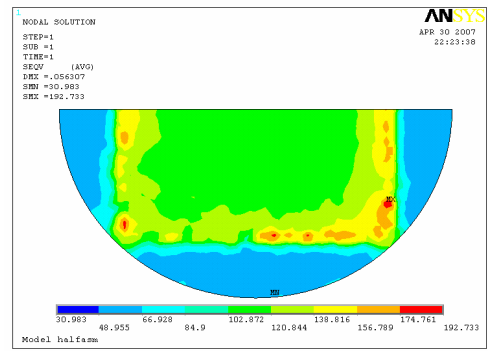

(b)

Fig. 6. Distributions of thermo-mechanical stress in LEDs with plastic packages: (a) stress in the chip, (b) stress in the die attachment.

## V. CONCLUSIONS

Thermal and mechanical analysis was done for the LEDs with ceramic packages. The obtained thermal and mechanical performance of LEDs with ceramic packages was compared with the LEDs with plastic packages. The ceramic packages showed better thermal performance than the plastic packages. The thermo-mechanical stress possibly induced in the mounting process was simulated using FEA method. Higher thermo-mechanical stress grew during the heating process for the ceramic packages despite of less mismatching CTEs. The high modulus of ceramic mould was proposed to be responsible. The results suggest that the mounting process of the high power LEDs with ceramic packages is critically important and should be in charge of delaminating interface layers in the package.

## ACKNOWLEDGMENT

This work was supported by the Korea Institute of Industry Technology Evaluation & Planning.

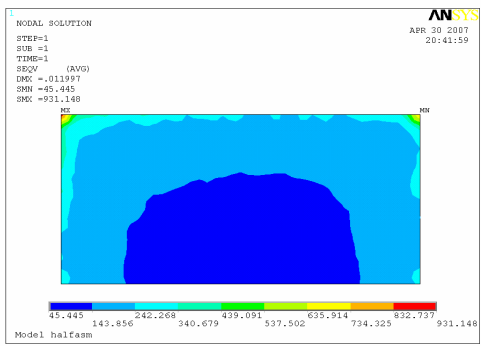

(a)

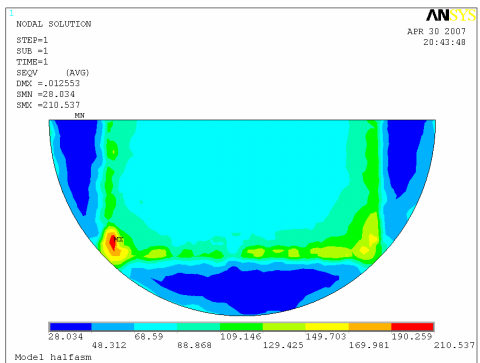

(b)

Fig. 5. Distributions of thermo-mechanical stress in LEDs with ceramic packages: (a) stress in the chip, (b) stress in the die attachment.